\documentclass{iaus}
\usepackage{graphicx}

\title[Period distribution and frequency of short-period binaries in the LMC]
{Analysis of the eclipsing binaries in the LMC discovered by
  OGLE: Period distribution and frequency of the short-period
  binaries}

\author[Mazeh, Tamuz \& North]   
{Tsevi Mazeh$^1$, Omer Tamuz$^1$ \and Pierre North$^2$}

\affiliation{
 $^1$School of Physics and Astronomy, Raymond and Beverly Sackler Faculty of Exact
Sciences, Tel Aviv University, Tel Aviv, Israel \break email: mazeh@wise.tau.ac.il\\
$^2$ Laboratoire d'Astrophysique, Ecole Polytechnique F\'ed\'erale de Lausanne
(EPFL), \\
Observatoire, CH-1290 Sauverny,
Switzerland}

\pubyear{2006}
\volume{IAUS240}  
\pagerange{1-10}
\date{?? and in revised form ??}
\jname{Proceedings Title IAU Symposium}
\editors{W. Hartkopf, E. Guinan \& P. Harmanec, eds.}
\begin{document}

\maketitle

\begin{abstract}

We review the results of our analysis of the OGLE LMC eclipsing
binaries (\cite[Mazeh, Tamuz \& North 2006]{mazeh06}), using EBAS ---
Eclipsing Binary Automated Solver, an automated algorithm to fit
lightcurves of eclipsing binaries (\cite[Tamuz, Mazeh \& North
2006]{tamuz06}).  

After being corrected for observational selection effects, the set of
detected eclipsing binaries yielded the period distribution
and the frequency of all LMC short-period binaries, and not just the
eclipsing systems.

Somewhat surprisingly, the period distribution is consistent with a
flat distribution in log P between 2 and 10 days.  The total number of
binaries with periods shorter than 10 days in the LMC was estimated
to be about 5000. This figure led us to suggest that $(0.7\pm 0.4)$\%
of the main-sequence A- and B-type stars are found in
binaries with periods shorter than 10 days. This frequency is
substantially smaller than the fraction of binaries found by small
Galactic radial-velocity surveys of B stars.

\end{abstract}

\firstsection 
              
\section{Introduction}

The OGLE project yielded a huge photometric dataset of the LMC
(\cite[Udalski \etal\ 2000]{udalski2000}), which includes a few
thousand eclipsing binary lightcurves (\cite[Wyrzykowski \etal\
2003]{lukas2003}). This dataset allows for the first time a
statistical analysis of the population of short-period binaries of an
entire galaxy. 

To analyse this set of lightcurves we constructed EBAS (Eclipsing
Binary Automated Solver), an automated algorithm (\cite[Tamuz, Mazeh
\& North 2006]{tamuz06}) to fit eclipsing lightcurves. Having solved
the lightcurves with EBAS, we proceeded to derive the statistical
features of the eclipsing binaries of the LMC (\cite[Mazeh, Tamuz \&
North 2006]{mazeh06}).  In this short paper we present the EBAS
algorithm and show two results of interest: a flat log-period
distribution and a lower-than-expected binary frequency of the
short-period early-type binaries.

\section{The EBAS Algorithm}
\label{sec:parameters}

EBAS is based on the EBOP code (\cite[Popper \& Etzel 1981, Etzel
1981]{PE81,etzel81}), which consists of two main components. The first
component generates a lightcurve for a given set of orbital elements
and stellar parameters, while the second finds the parameters that
best fit the observational data. We only used the lightcurve
generator, and wrote our own code to search for the best-fit set of
elements that minimize the  $\chi^2$ statistic.

The search for the global $\chi^2$ minimum is performed in two
stages. We first find a good initial guess, and then use a simulated
annealing algorithm to find the global minimum.  While the first stage
is merely aimed at finding an initial guess for the next stage, in
most cases it already converges to a very good solution. A full description
of the algorithm is given in \cite{tamuz06}.

During the development of EBAS we found that some solutions with low
$\chi^2$ might be unsatisfactory. While, for such solutions, the value
of $\chi^2$ is reasonable, a visual inspection of the residuals,
plotted as a function of phase, revealed that the fit is
sub-optimal. For such cases, human interaction was needed
to improve the fit, or to otherwise decree the solution
unsatisfactory.  In order to allow an automated approach, an automatic
algorithm must replace human evaluation.

We therefore defined a new estimator which is sensitive to the
correlation between adjacent residuals of the measurements relative to
the model. This feature is in contrast to the behaviour of the
$\chi^2$ function, which measures the sum of the squares of the
residuals, but is not sensitive to the signs of the different
residuals and their order.

Denoting by $k_i$ the number of residuals in the $i$-th run (a series
of consecutive residuals with the same sign), we defined
the 'alarm' $\mathcal A$ as:

\begin{equation}
\mathcal{A}=\frac{1}{\chi^2}\sum_{i=1}^M{\left(\frac{r_{i,1}}{\sigma_{i,1}}
                                  + \frac{r_{i,2}}{\sigma_{i,2}}
                                  +\cdots+
                                  \frac{r_{i,k_i}}{\sigma_{i,k_i}}
                                  \right)^2}
                                  -(1+\frac{4}{\pi})\ \ ,
\end{equation}
where $r_{i,j}$ is the residual of the $j$-th measurement of the
$i$-th run and $\sigma_{i,j}$ is its uncertainty. The sum is over all
the measurements in a run and then over the $M$ runs.
Dividing by $\chi^2$
assures that, in contrast to $\chi^2$ itself, $\mathcal A$ is not
sensitive to a systematic overestimation or underestimation of the
uncertainties.
It is easy to see that $\mathcal A$ is minimal when the residuals
alternate between positive and negative values, and that long runs
with large residuals increase its value. 

When a solution found by EBAS showed high $\mathcal A$, EBAS
automatically classified the solution as unsatisfactory and started a
modified search of the parameter space in order 
to find a better solution. Systems for which a low enough
$\mathcal A$ solution could not be found were marked as such. Visual
inspection of these lightcurves and their fitted models showed that most
of them are close contact systems which EBOP did not model well.

\section{Analysis of the OGLE LMC Eclipsing Binaries}   
\label{elements}

The LMC OGLE-II photometric campaign (\cite[Udalski \etal\
2000]{udalski2000}) was carried out from 1997 to 2000, during which
between 260 and 512 measurements in the $I$ band were taken for 21
fields (\cite[Zebrun \etal\ 2001]{zebrun2001}).  \cite{lukas2003}
searched the photometric data base and identified 2580 binaries. We
analysed those systems with EBAS and found 1931 acceptable
solutions. We excluded all binaries with sum of radii larger than 0.6
of the binary separation. Furthermore, we excluded also binaries with
high $\mathcal A$ values, which indicated that they are probably
contact systems that EBOP can not model properly (see \cite{mazeh06}
for details).

\begin{figure*}
 \includegraphics{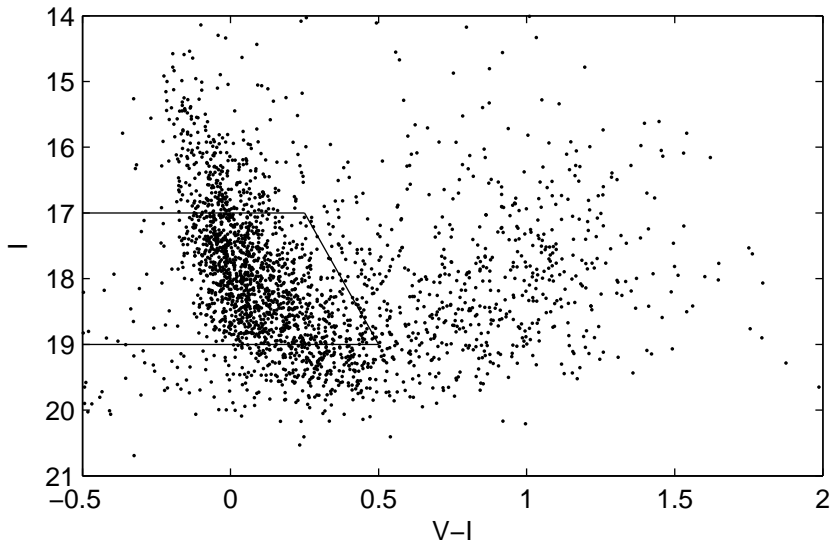}
 \caption{The $I$-magnitude as a function of the averaged $V-I$ colour
 of all eclipsing binaries of the sample. Only binaries that appear
 inside the trapezoid were included in the trimmed sample.}
 \label{fig:VminusI}
\end{figure*}

The very large sample of 1931 short-period binaries enables us to
derive some statistical features of the population of short-period
binaries in the LMC. However, the sample suffers from serious
observational selection effects, which affected the discovery of the
eclipsing binaries. To be able to correct for the selection effects we
needed a well-defined homogeneous sample. We therefore trimmed the
sample before deriving the period distribution. Our trimmed sample
consisted of all systems of magnitude between 17 and 19 in the $I$
band with periods shorter than 10 days and having main-sequence color
(see Fig~\ref{fig:VminusI}).  In this way we chose only systems
with A and B main-sequence primaries.  See~\cite{mazeh06} for a full
description of the trimmed sample.

\subsection{The period distribution}

In order to explore the period distribution of the LMC short-period
binaries, and not just the eclipsing binaries, we needed to correct
for the selection effects which affect the likelihood of a binary
system to be eclipsing, and to be detected by the OGLE survey. We did
this by calculating for each eclipsing binary in the trimmed sample a
weight $w$, which was the reciprocal of the probability of being
detected as an eclipsing binary, assuming random orientation and phase
(see \cite[Tamuz, Mazeh \& North 2006]{tamuz06}). When calculating
the period distribution of the binary population we then considered that
eclipsing binary as $w$ systems.

\begin{figure*}
 \includegraphics{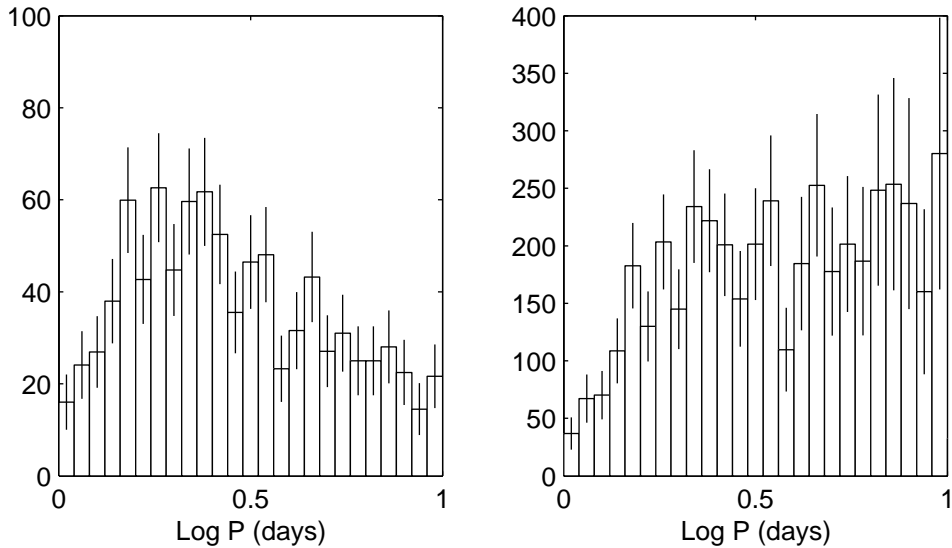}
 \caption{The period distribution of the binaries in the LMC. The left
 panel shows the period histogram of the trimmed sample, while the
 right one shows the corrected histogram}
 \label{fig:period_distr}
\end{figure*}

We plotted in Fig~\ref{fig:period_distr} two period histograms of the
trimmed sample, before and after the correction for the observational
effects was applied. We emphasize that if the correction was applied
properly, the right panel represents the period distribution of all
binaries in the LMC with $I$-magnitude between 17 and 19, and with sum
of radii smaller than 0.6, and not only the eclipsing binaries.

The corrected period histogram shows clearly a distribution that rises
up to about $\log P=0.3$, and then flattens off, suggesting that the
period distribution of the short-period binaries with early-type
binaries is consistent with {\it a flat log distribution between 2 and
10 days}.

\subsection{The frequency of the short-period binaries}

As stated above, the corrections applied by \cite[Mazeh, Tamuz \& North
2006]{mazeh06} allowed us to study the
distributions of all short-period binaries, up to 10 days, and not
only the eclipsing binaries. We wished to take advantage of this feature
of our analysis and estimate, within the limits of the sample, the
total number of binaries in the LMC and their fractional frequency.

The sum of weights of the 938 binaries in the trimmed sample was
4585. This means that to our best estimate there are 4585 short-period
binaries in the LMC that fulfill the constraints we had on the trimmed
sample. To get the total number of binaries with periods shorter than
10 days we had to add the high-alarm binaries and the ones with large
radii; since the added systems are close binaries, their detection
probability reaches almost unity so we did not correct for undetected
binaries in their case. When we added those we ended up with 5004
binaries.

We therefore suggested that the number of binaries in the LMC, with
period shorter than 10 days, with $I$ between 17 and 19, and for which
the $V-I$ colour of the system indicates a main-sequence primary, is
about 5000. Assuming \cite{lukas2003} might not have detected all the
eclipsing binaries in the dataset and classified them as such, we
arbitrarily assigned an error of 30\% to the number of binaries in the
sample.

In order to estimate the fractional frequency of A- and B-type
main-sequence stars which reside in binaries that could have been
detected by OGLE we had to estimate how many main-sequence {\it
single} stars were found by OGLE in the same range of magnitudes and
colours. To do that, we applied exactly the same trimming procedure we
performed above to the whole OGLE dataset of LMC stars and found
332,297 main-sequence stars in the $I$-range of 17--19. However, the
luminosity of a binary is brighter than the luminosity of its primary
by a magnitude that depends on the light ratio of the two stars of the
system. Therefore, we considered stellar range of $I$ = 17.5 - 19.5,
and found 705,535 stars. We therefore adopted 700,000 as a
representative number of single stars in the LMC similar to the
population of our binary sample. We arbitrarily assign an error of
50\% to this figure.

We therefore concluded that $(0.7\pm 0.4)$\% of the main-sequence A-
and B-type stars in the LMC are found in binaries with periods shorter
than 10 days.

\section{Discussion}            %
\label{discussion}

\subsection{The period distribution}

The log flat period distribution derived here is probably not
consistent with the period distribution of \cite{DM91}, who adopted a
Gaussian with $\overline{\log P}=4.8$ and $\sigma=2.3$, $P$ being
measured in days.  Their distribution would provide within the same log P
interval almost twice as many systems at $\log P = 1.0$ than at $\log
P = 0.3$ (more precisely, the factor would be 1.7), while the new
derived distribution is probably flat.

It is interesting to note that \cite{heacox1998} reanalysed the data
of \cite{DM91} and claimed that $f(a)$, the distribution of the
G-dwarf orbital semi-major axis, $a$, is $f(a) \propto a^{-1}da$,
which implies a flat log orbital separation distribution. This is
equivalent to the present probable result, although the latter refers
to LMC binaries with A- and B-type primaries, and is limited only to a
very small range of orbital separation.

The flat log-period distribution implies a flat log orbital-separation
distribution for a given total binary mass. Such a distribution
indicates that there is no preferred length scale for the formation of
short-period binaries (\cite[Heacox 1998]{heacox1998}), at least in
the range between 0.05 and 0.16 AU, for a total binary mass of 5
$M_{\odot}$. Alternatively, the results indicate that the specific
angular momentum distribution is flat on a log scale at the $10^{19}$
$cm^2s^{-1}$ range.

\subsection{The binary fraction}

The fractional frequency found here is surprisingly small --- $(0.7\pm
0.4)$\%. It would be interesting to compare the frequency of B-star
binaries found here for the LMC with a similar frequency study of B
stars in our Galaxy. However, such a large systematic study of
eclipsing binaries is not available. Instead, a few radial-velocity
and photometric searches for binaries in relatively small Galactic
samples were performed. \cite{wolff78} studied the frequency of
binaries in sharp-lined B7--B9 stars.  Her Table~1 presents 73 such
stars with $V\sin i < 100$~km\,s$^{-1}$, among which 17 are members of
binary systems with $P_{\mathrm orb} < 10$~days. Her sample yields a
percentage of $23$\%, one and a half orders of magnitude larger than 
what we found in the LMC.

Similar and even higher values of binary frequency were derived for B
stars in Galactic clusters and associations.  \cite{ML91} have
determined the rate of binaries among B stars in the Orion OB1
association, and obtained an overall frequency of binaries with
$P_{\mathrm orb} < 10$~days of $26\pm 6$\%. \cite{GM01} found a rate
of binaries as high as $82$\% among stars hotter than B1.5 in the very
young open cluster NGC~6231, though in a sample of 34 stars only. Most
orbits have periods shorter than 10 days.  \cite{R96} estimated a rate
of binaries of at least $52$\% for 36 B1-B9 stars in the same cluster.

These studies show that binaries hosting B-type stars tend to be
more frequent in young Galactic clusters than in the field. However,
even though our LMC binaries are representative of the LMC field
rather than of LMC clusters, the frequency we derived appears much
lower than that of the binaries in the Galactic field.

A few photometric studies were devoted to the search of eclipsing
binaries in Galactic globular clusters (e.g., \cite[Yan \& Reid 1996,
Kaluzny 1997]{yan96, kaluzny97}). One such study was the HST 8.3-day
observations of 47 Tuc (\cite[Albrow \etal\ 2001]{albrow01}), which
monitored 46,422 stars and discovered 5 eclipsing binaries with
periods longer than about 4 days. From their conclusion we derive
$(2\pm1)$\% for periods between 2 and 10 days. This value, derived for
late-type stars, is much smaller than the frequency derived by the
radial-velocity surveys of B stars, and is close to the
$(0.7\pm0.4)$\% frequency we derived for the LMC.

We conclude, therefore, that the frequency of binaries we found in the
LMC is substantially smaller than the frequency found by Galactic
radial-velocity surveys. The detected frequency of B-type binaries is
larger by a factor of 30 or more than the frequency we found, while
the binary frequency of K- and G-type stars is probably larger by a
factor of four (\cite{HMUA03} found $2.7\pm 0.8$ per cent binaries
with $P< 10$~days in the solar neighbourhood). On the other hand, the
binary frequency found by photometric searches in 47 Tuc is only
slightly higher and still consistent with the frequency we deduced for
the LMC. It seems that the frequency derived from photometric searches
is consistently smaller than the one found by radial-velocity
observations. We are not aware of any observational effect that could
cause such a large difference between the radial velocity and the
photometric studies. Therefore, we suggest that the large difference
in the binary frequencies is probably real and remains a
mystery. Obviously, it would be extremely useful and interesting to
have studies of eclipsing binaries similar to the present one in our
Galaxy, as well as in other nearby galaxies.

\section*{Acknowledgments}
We are grateful to the OGLE team, and to L. Wyrzykowski in particular,
for the photometric data set and for the eclipsing binary
analysis. This work was supported by the Israeli Science Foundation
through grant no. 03/233.

\end{document}